\documentclass[twocolumn,superscriptaddress,aps,prl]{revtex4-1}

\usepackage{tabularx}
\usepackage{amsmath}
\usepackage{amssymb}
\usepackage{bm}
\usepackage{graphicx}
\usepackage{psfrag}
\usepackage{url}

\begin{document}

\title{Theory of wetting-induced fluid entrainment by advancing contact lines on dry surfaces} 
\author{\firstname{R.} \surname{Ledesma-Aguilar}}
\email{r.ledesmaaguilar1@physics.ox.ac.uk}
\affiliation{The Rudolf Peierls Centre for Theoretical Physics,  
University of Oxford, 1 Keble Road, Oxford OX1 3NP, United Kingdom}
\author{\firstname{A.} \surname{Hern\'andez-Machado}}
\affiliation{Departament d'Estructura i Constituents de la
  Mat\`eria. Universitat de Barcelona, C. Mart\'i i Franqu\`es 1, E-08028 Barcelona, Spain}
\author{\firstname{I.} \surname{Pagonabarraga}}
\affiliation{Departament de F\'isica Fonamental.  Universitat de
  Barcelona, C. Mart\'i i Franqu\`es 1, E-08028 Barcelona, Spain}  
\date{\today}

\begin{abstract}
We report on the onset of fluid entrainment when a contact line is forced to advance over a dry solid of arbitrary wettability.
We show that entrainment occurs at a critical advancing speed beyond which the balance between capillary, viscous and contact line forces 
sustaining the shape of the interface is no longer satisfied.      
Wetting couples to the hydrodynamics by setting both the morphology of the interface 
at small scales and the viscous friction  of the front.  
We find that the critical deformation that the interface can sustain is controlled by the friction at the contact line and the viscosity contrast between the displacing 
and displaced fluids, leading to a rich variety of wetting-entrainment regimes.  
We discuss the potential use of our theory to measure contact-line forces using atomic force microscopy, and to study entrainment 
under microfluidic conditions exploiting colloid-polymer fluids of ultra-low surface tension. 
\end{abstract}

\maketitle

\newpage

{\it Introduction.--} 
The entrainment of a fluid by a solid raises fundamental questions on the hydrodynamics of moving
contact lines~\cite{Rolley01}, which are the intersection of fluid and solid boundaries, and is relevant to a wide variety of practical situations 
in materials science~\cite{Pagonabarraga03,Subramaniam-NatMater-2006} and microfluidics~\cite{Utada-PhysRevLett-2007,Leshansky-PhysRevLett-2012}.
Fluid entrainment gives rise to rich  phenomena such as bubble entrainment in 
solid plate immersion~\cite{Blake-Nature-1979,Marchand-PhysRevLett-2012}, gas intrusion in coating~\cite{Blake-AdvCollIntSci-2002,*BlakePatent} and solid-liquid splashing~\cite{Duez01}, 
drop emission from  forced liquid  microfilaments~\cite{Ledesma-NatMat-09}, the ejection of drops and rivulets from forced running drops~\cite{Podgorski-PRL-2001,Snoeijer-PRL-09} and
the more familiar film deposition on withdrawn solid plates~\cite{Maleki-Langmuir-2007,Snoeijer-PRL-08}.   

Typically, entrainment has been studied when a liquid front is forced to retreat from a solid at constant driving speed, $U$. 
Considerable theoretical progress has been made in understanding the stability of such {\it receding} configuration on hydrophilic substrates \cite{deGennes03,Eggers01}, where the static contact angle, or Young's angle $\theta_{\rm e}$, is small.  
In this case, it is now well understood that the front can only dewet completely from the surface up to a maximum receding speed~$U_{\rm{rec}} \sim A \gamma \theta_{\rm e}^3 / \eta_{\rm L}$~\cite{deGennes03,Eggers01}, where $\gamma$  is the liquid/solid interfacial surface tension,  $\eta_{\rm L}$ is the liquid viscosity, and  $A$ is a numerical prefactor specific to the system  geometry.  Above this threshold, the liquid
is entrained by the solid and a thin film is left on the surface~\cite{Snoeijer-PRL-08}. 

Despite being an archetype of forced liquid fronts, the {\it advancing} configuration, where a liquid moves over a solid surface displacing 
a gas, is understood to a much lower extent. For advancing contact lines, the front is also destabilized above  a critical speed, $U_{\rm{adv}}$.   However,  this 
can be significantly larger ($\sim$ m~s$^{-1}$) than the receding speed ($\sim {\rm cm~s^{-1}}$)~\cite{Snoeijer-PRL-09}.  
For a receding contact line the liquid dewets from the solid, while for an advancing contact line it is the gas.  This asymmetry
can be used to rationalize the gap in magnitudes between $U_{\rm adv}$ and $U_{\rm rec}$;
upon retreating, the gas offers a smaller viscous friction than the liquid,  and thus the destabilization threshold should be larger for the advancing 
configuration. Although  a similar scaling,  $U_{\rm{adv}} \sim B \gamma (\pi - \theta_{\rm e})^3/\eta_{\rm L}$ \cite{Duez01}, has been suggested
 for advancing contact lines on hydrophobic surfaces  (where $B\gg A$  depends on the gas viscosity, $\eta_{\rm G}$), this relation does not capture the experimentally reported  dependence 
 of $U_{\rm adv}$ on the wider $\theta_{\rm e}$ range~\cite{Blake-AdvCollIntSci-2002,Duez01,Ledesma-NatMat-09}.

{ Increasing experimental evidence supports the important role of surface specificity on entrainment~\cite{Blake-AdvCollIntSci-2002,Duez01,Ledesma-NatMat-09}.}  
A non-monotonic dependence of $U_{\rm adv}$ as $\theta_{\rm e}$ is varied was observed when forcing water films on wafers covered with 
dry gelatin containing different surfactants~\cite{Blake-AdvCollIntSci-2002}. 
However, careful measurements displayed a dramatic saturation of $U_{\rm{adv}}$ on hydrophilic substrates made of glass, and a rapid decay 
with $\theta_{\rm e}$ on hydrophobic solids (treated with silane chains)~\cite{Duez01}. Similarly, $U_{\rm adv}$ can be controlled by wearing a superhydrophobic surface, suggesting that
surface heterogeneity can lead to entrainment~\cite{Ledesma-NatMat-09}. 
These examples illustrate the wide variety of situations where entrainment can arise, yet for which a general theoretical 
framework is lacking.

In this Letter we put forward a theoretical framework that combines the hydrodynamics of the advancing front with the dynamics 
of the contact line to predict the onset of fluid entrainment on surfaces of arbitrary wetting properties.     
Due to the strong sensitivity of the advancing front to   the interfacial morphology, the maximum advancing speed of the front shows a rich behavior 
depending on the wetting properties of the solid and the viscosity contrast between the fluids.
The coupling with the large-scale morphology of the front in our
theoretical framework is generic; hence the reported results can 
be applied to a wide variety of fluid geometries.

\begin{figure}[t!]
\includegraphics[width=0.35\textwidth]{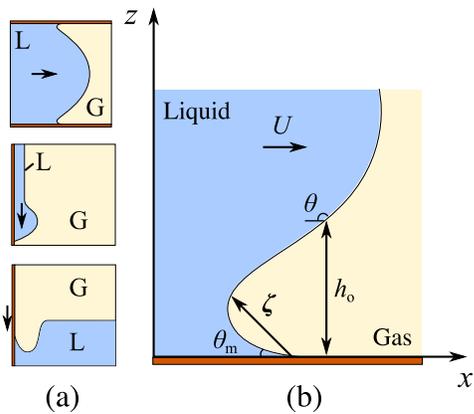}
\caption{(a) Schematic of typical advancing liquid fronts: a meniscus is forced between parallel plates (top), a thin liquid film
runs down a vertical surface (middle) and a solid plate plunges into a liquid bath (bottom). The arrow shows the 
direction of the imposed driving.
(b) Below the cross-over length $h_{\rm o}$ the profile
is determined by capillary and viscous forces, and is characterized by the dynamic and microscopic angles, $\theta_{\rm }$ and 
$\theta_{\rm m}$.  
\label{fig:Scheme}}
\end{figure}

{\it Theory.--} Fig.~\ref{fig:Scheme}(a) shows  schematically  typical liquid fronts advancing over dry substrates due to the 
action of different external forces, ${\bf F}_{\rm{ext}}$, {\it e.g.}, a pressure gradient forcing a meniscus between parallel plates, 
gravity pushing a thin film  down a vertical surface, or the drag caused by a solid plate that plunges into a liquid bath.   
At small scales the shape of the front is independent of geometry, and has the generic structure 
depicted in Fig.~\ref{fig:Scheme}(b).  
Here we consider the motion of flat contact lines and focus on the variation of 
the front in the $xz$ plane, with the solid surface located at  $z=0$. 
In steady state, the interface profile is described by the parametric curve ${\boldsymbol \zeta} (s) = f(s)\hat{\bf e}_x + h(s)\hat{\bf e}_z$ which propagates at speed $U$. 
The relevant condition for a steady interfacial state is that the total force per unit length of the contact line acting along the longitudinal 
coordinate, $x$, vanishes, 
\begin{equation} 
F_x(\boldsymbol \zeta ,U)= F_{\rm ext} + F_{\gamma} +F_{\eta_{\rm L} }+F_{\eta_{\rm G}}=0,
\label{eq:ForceBalance1}
\end{equation}
where $F_\gamma$ is the capillary force, and $F_{\eta_{\rm L}}$ and $F_{\eta_{\rm G}}$ are the viscous friction forces offered 
by the liquid and the gas. Since these forces depend on $\boldsymbol \zeta$ and $U$, it is possible to recast 
Eq.~(\ref{eq:ForceBalance1}) into a relationship between the interface shape and its velocity.
Hence, the entrainment onset corresponds to the maximum speed for which the interface shape is consistent with the previous force balance.

Obtaining the maximum advancing speed, $U_{\rm adv}$, 
is challenging because the interfacial shape follows  from the solution of the non-linear free-boundary problem 
associated with the hydrodynamics of the liquid and gas phases, 
subject to boundary conditions at the solid surface.  While numerically it is possible to solve the hydrodynamics for a specific geometry~\cite{Marchand-PhysRevLett-2012},  this gives a less general understanding of the physics behind fluid entrainment. As an alternative, we will show that a good approximation of the curved interface profile captures  the leading order behavior of the force balance.  

\begin{figure}[t!]
\centering
\includegraphics[width=0.45\textwidth]{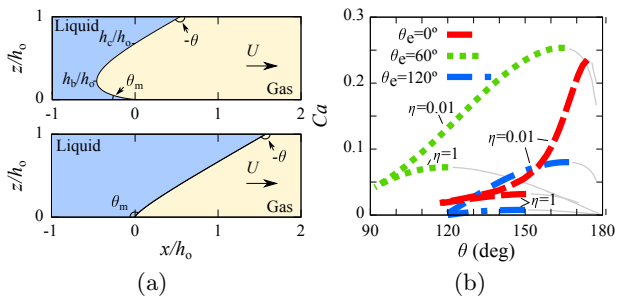}\\
(a)\hspace{3.8cm}(b) 
\caption{
(a) Predicted inner interface profiles 
on hydrophilic ($\theta_{\rm m} = 15^\circ$, top) and hydrophobic ($\theta_{\rm m}=120^\circ$, bottom) surfaces (located at $z=0$) for  
$\theta = 150^\circ$ and $\xi/h_{\rm o}=10^{-2}$.
(b) Relation between the capillary number and the dynamic angle at different viscosity contrasts for steady interface profiles 
for $\xi/h_{\rm o} = 10^{-5}$, $c_{\rm L}=c_{\rm G}=4.0$, and $w_{\rm m}=1.0$.  Above the maxima the inner-region force balance is no longer satisfied, leading to entrainment. 
Thin lines to the right of maxima correspond to unstable solutions of the force balance.
\label{fig:shapes_angles}}
\end{figure}

The scale dependence of the competing terms in Eq.~(\ref{eq:ForceBalance1}) gives rise to a natural division of the interface into two regions. At large scales, corresponding to an outer region, the 
interface shape is determined by the balance between the external forcing and capillarity (see Fig.~\ref{fig:Scheme}(a)). 
This contrasts with a small-scale inner region, depicted in Fig.~\ref{fig:Scheme}(b), where viscous stresses and capillarity are 
dominant. 
The cross-over between the outer and inner regions occurs at a thickness $h_{\rm o}$, comparable to  
the capillary length
$\ell_{\rm c} \equiv  \sqrt{\gamma/\partial_x P_{\rm ext}}$, { where $\partial_x P_{\rm ext}$ is the hydrostatic pressure gradient due to the external force.} 
Since the front is driven externally, the stability of the interface is subject to the ability of the contact line to follow the leading front. 
Accordingly, we focus on the dynamics in the region close to the solid substrate.  In the inner region the external force can be neglected while  the capillary 
term corresponds to the dynamic Young's force, $F_\gamma = \gamma(\cos\theta_{\rm m} - \cos{\theta_l} + \kappa h_{\rm o})$, obtained integrating the gradient of the Laplace pressure, 
$\gamma \kappa'$, for $\xi \leq h \leq h_{\rm o}$, where $\kappa$ and $\theta_l$ are the the local curvature and inclination angle of the interface,  $\xi$ is the molecular thickness, 
and $\theta_{\rm m}$ is the microscopic contact angle with which the interface intersects the solid. 
{The remaining terms in Eq.~(\ref{eq:ForceBalance1}) are the friction forces},  $F_{\eta_i} = - 3 \eta_i U \int_{\Omega_i}  \frac{\mathrm d\Omega \mathcal F(h')}{h^2} - c_i \eta_iU,$
with $i=\{\mathrm G,L\}$, where $\Omega_i$ refers to the area occupied by phase $i$ in the $xz$ plane (Fig.~\ref{fig:shapes_angles}(a)). 
The first term in $F_{\eta{_i}}$ accounts for the friction arising from the sliding motion of the fluid wedges meeting at the contact line, where 
the large-slope correction, $\mathcal F(h')$, approaches unity for sharp wedges~\cite{Snoeijer01}, close to the entrainment threshold. 
{The second term corresponds to viscous stresses arising far from the contact line and is characterized by order-unity numerical pre-factors, $c_i$.}
Dividing through by $\gamma$, Eq.~(1) reduces to 
\begin{multline}\label{eq:ForceBalance0}
\left\{\cos\theta_{\rm m} - \cos{\theta_l}+ \kappa h_{\rm o}\right\} - \left\{\frac{c_{\rm L}}{3} + \int_{\Omega_{\rm L}}  \frac{\mathrm d\Omega \mathcal F(h')}{h^2} \right\}3Ca\\ 
- \eta\left\{\frac{c_{\rm G}}{3} +  \int_{\Omega_{\rm G}}  \frac{\mathrm d\Omega \mathcal F(h')}{h^2} \right \}3 Ca = 0,
\end{multline}
where  $Ca\equiv \eta_{\rm L}U/\gamma$ is the capillary number and $\eta \equiv \eta_{\rm G}/\eta_{\rm L}$ is the viscosity contrast between the fluids.
This equation includes the effects of both moving phases, and can be used to describe advancing and receding fronts.

Previous results for receding contact lines correspond to the limit of small interface slopes~\cite{Eggers01}, and vanishing
gas viscosity. 
This limit is recovered by setting $c_{\rm L}=0$ and $\eta = 0$ in Eq.~(\ref{eq:ForceBalance0}), whereby one obtains
$(\theta^2_{\rm m}-\theta_l^2)/2 + h'' h_{\rm o}\approx 3 Ca \int_{\Omega_{\rm L}}  \frac{\mathrm d\Omega}{h^2}$. This expression is equivalent
to the thin-film lubrication equation, as pointed out in Ref.~\cite{Rolley01}, which can be used to obtain the well-known scaling for the maximum receding 
speed $U_{\rm rec} \sim A\gamma \theta_{\rm e}^3/\eta_{\rm L}$~\cite{Eggers01}.

The first challenge in using Eq.~(\ref{eq:ForceBalance0}) 
is the divergence of viscous forces at the intersection between 
the fluid and solid boundaries~\cite{Huh01}.  To regularize this singularity we treat $\xi$ as a cut-off length scale. Hence, at  the 
solid boundary the profile obeys 
\begin{equation}
\label{eq:MC2}
h=\xi\qquad {\rm and} \qquad h'(h=\xi)=\tan\theta_{\rm m},
\end{equation}
where $\theta_{\rm m}$ determines  the local structure of the interface at small scales.
While for receding contact lines  $\theta_{\rm m} \approx \theta_{\rm e} \ll 1$~\cite{Cox-JFluidMech-1986}, 
for advancing fronts the interface slope can deviate from its static value for arbitrarily small distances from the solid surface due to the microscopic details of the contact 
line motion~\cite{deGennes03}.  We  account for the detailed dynamics at the contact line assuming a local force balance, which to leading order in the front displacement 
can be expressed as~\cite{Rolley01}  
\begin{equation}
\cos \theta_{\rm m} = \cos \theta_{\rm e} + w_{\rm m} Ca,
\label{eq:CLfriction}
\end{equation} 
where $w_{\rm m}$ is a dimensionless friction coefficient that subsumes the specific microscopic details of the contact line motion~\cite{Yeomans01,Qian01}.  
For advancing fronts the interface curvature changes sign at  the cross-over thickness, $h_{\rm o}\sim \ell_{\rm c}$, due to the external forcing, 
and its local slope is characterized by the 
dynamic angle, $\theta$. The matching conditions of the inner region with the outer profile hence read
\begin{equation}
\label{eq:MC1}
h' (h = h_{\rm o})= \tan\theta \quad {\rm and} \qquad \kappa (h = h_{\rm o}) = 0,
\end{equation}
where $h'$ is the local interfacial slope.   
Rather than using these conditions to relate  the inner and outer profiles of a specific geometry, we consider a generic {setting} 
by treating $\theta$ as a given external interfacial deformation. 

To determine the interfacial profile, we use the  local force-density balance $\kappa' \sim g(Ca)/h^2 $, where the local capillary force, varying as $\kappa \approx h''$, is balanced by the viscous force, scaling as $h^{-2}$. 
Since for advancing fronts the profile may develop overhangs, it is useful  to invert the profile, $ f(h) \equiv x = \theta_{\rm m}^{-1}h + g(Ca) f_1(h) + \cdots$, for small $g(Ca)$. Matching $f(h)$ with the outer region through  Eq.~(\ref{eq:MC1}), we obtain both the interface profile
\begin{equation}
f(h) \approx \frac{h-\xi}{\tan\theta_{\rm m}} + g(Ca)\left[h\ln \frac{h}{\xi}-h+\xi-\frac{(h-\xi)^2}{2h_{\rm o}}\right]
\label{eq:profile}
\end{equation}
and  $g(Ca) =  (\tan\theta_{\rm m}-\tan\theta)/\{\tan\theta_{\rm m}\tan \theta(\ln(h_{\rm o}/\xi)-1+\xi/h_{\rm o})\}$~\footnote{See supplementary material for details}.  
For small interfacial slopes $g(Ca) \rightarrow 3Ca$~\footnote{This particular limit corresponds to the lubrication equation, which can be solved perturbatively in powers of $Ca$.}. 
Expanding $h$ in powers of $Ca$, together with Eq.~(\ref{eq:MC2}) to fix the profile at the contact line,  one recovers the classic result $h(x,Ca)=\theta_{\rm m} x+ 3 Ca h_1(x) + {\cal O}(Ca^2),$ where $h_1\sim \ln(x/\xi)$~\cite{Cox-JFluidMech-1986}.

{The wettability of the solid surface has a strong influence on the front morphology.
To illustrate this we plot Eq.~(\ref{eq:profile}) in Fig.~\ref{fig:shapes_angles}(a)
for hydrophilic ($\theta_{\rm m} = 15^\circ$) and hydrophobic ($\theta_{\rm m} = 120^\circ$) surfaces for a fixed  dynamic contact angle, $\theta = 150^\circ$. We fix the scale separation as $\xi/h_{\rm o}  = 10^{-2}$
for visualisation purposes, although  
{such weak separation between microscopic and macroscopic lengthscales
is realistic, {\it e.g.}, for colloid-polymer demixed fluids~\cite{Aarts01}}.
Due to the coupling to the solid at small scales, Eq.~({\ref{eq:MC2}}), the interface can bend forward to develop
a foot on hydrophilic substrates (top panel). This structure appears
whenever $\theta_{\rm m} \leq 90^\circ$ and extends from the molecular length, $\xi$, to the turning point $h=h_{\rm b}$.   
The gas counterpart to the liquid foot is the wedge shown in Fig.~\ref{fig:shapes_angles}(a), which can only form when $\theta > 90^\circ$.  On hydrophilic substrates, when $\theta_{\rm m} \leq 90^\circ$, 
the wedge is truncated and extends down to the thickness of the overhang at the contact line, $h_{\rm c}$ (top panel), or down to the microscopic lengthscale $\xi$ for hydrophobic surfaces (bottom panel).  
Remarkably, Eq.~(\ref{eq:profile}) is a very good approximation to the interface profile even for large slopes, 
as shown by a direct comparison to lattice-Boltzmann simulations~\footnote{See supplement for details on the geometry of the interface (the size of the foot, $h_{\rm b}$ and the depth of the gas wedge $h_{\rm c}$), and details
of the comparison to simulations}.}

With the shape of the interface, Eq.~(\ref{eq:profile}), we can now evaluate Eq.~(\ref{eq:ForceBalance0}) to obtain 
\begin{multline}\label{eq:ForceBalance}
\left\{\cos\theta_{\rm m} - \cos{\theta}\right\} - \left\{\frac{c_{\rm L}}{3} + H(h_{\rm b}) - H(\xi)\right\}3Ca\\ 
- \eta\left\{\frac{c_{\rm G}}{3} +  H(h_{\rm c}) - H(h_{\rm o})\right\}3 Ca = 0,
\end{multline}
where  $H(h)\equiv \ln(h)/\tan\theta_{\rm m } + g(Ca)\left[\ln^2(h/\xi)/2 - h/h_{\rm o} + (\xi/h_{\rm o})\ln(h)\right]$.  Note that, 
due to the definition of $\theta$ in Eq.~(\ref{eq:MC1}), the curvature dependent term drops in Eq.~(\ref{eq:ForceBalance}). 

\begin{figure}[t!]
\includegraphics[width=0.45\textwidth]{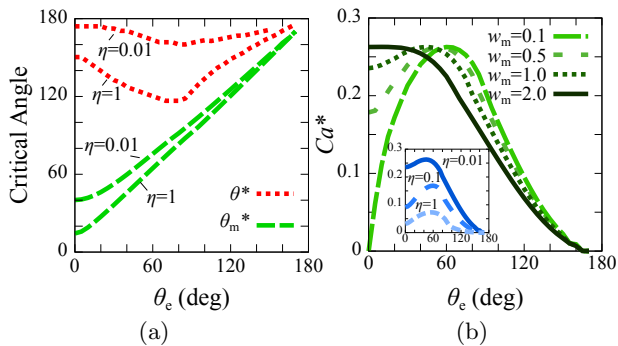}\\ 
(a)\hspace{3.8cm}(b) 
\caption{(a) Critical dynamic and microscopic angles as a function of the static angle. (b) Critical capillary number as a function of the static angle for different values of the contact line friction coefficient,  
and viscosity contrast (inset). 
All other parameter values are as in Fig.~\ref{fig:shapes_angles}(b).  \label{fig:CriticalCa}}
\end{figure}

{\it Discussion.--} {
Eq.~(\ref{eq:ForceBalance}) determines the velocity of the liquid front as a function 
of the dynamic angle over the wide range imposed by the external forcing.}
As depicted in Fig.~\ref{fig:shapes_angles}(b), $\theta$ increases from its  equilibrium value with $Ca$, 
indicating the morphological response of the interface to larger driving forces.
Crucially, the capillary force in Eq.~(\ref{eq:ForceBalance}) can only sustain a maximum deformation, while the friction forces can grow indefinitely
as $\theta\rightarrow 180^\circ$. 
This implies that there is a maximum capillary number, $Ca^*$, beyond which the force balance cannot be satisfied, leading to
fluid entrainment. 
The set of maxima shown in Fig.~\ref{fig:shapes_angles}(b) correspond to different critical
interface deformations, and therefore depend on the small-scale coupling set by $\theta_{\rm e}$ and on the viscous
bending, controlled by $\eta$.  
This is illustrated in Fig.~\ref{fig:CriticalCa}(a), which shows that $Ca^*$ is reached at a  critical dynamic angle, $\theta^*$, weakly dependent on $\theta_{\rm e}$ and is generally above $160^\circ$ for $ \eta=10^{-2}$ (representative of air-water systems). This changes for high $ \eta$, where a stronger viscous friction offered by the displaced phase leads
to lower $Ca^*$ and $\theta^*$.  Similarly, the microscopic contact angle reaches a critical value, $\theta^*_{\rm m}$, at the onset of entrainment, which differs from the equilibrium angle 
by an amount set by the viscosity contrast. 

Remarkably, the maximum advancing speed, $U_{\rm adv} = \gamma Ca^*/\eta_{\rm L} $, depends on the static angle {\em and} on the contact line friction~(Fig.~\ref{fig:CriticalCa}(b)).
On hydrophilic surfaces the friction due to the liquid foot controls $Ca^*$. 
For small $w_{\rm m}$ Eq.~(\ref{eq:CLfriction}) yields $\theta_{\rm m} \approx \theta_{\rm e}$, leading to a sharp foot that generates a larger amount of viscous friction. This increase in the viscous response becomes larger
as  $\theta_{\rm e}\rightarrow 0$, and therefore generates unstable fronts at smaller $Ca^*$. The result is a non-monotonic $Ca^*-\theta_{\rm e}$ dependence.
For larger $w_{\rm m}$, the microscopic angle departs from $\theta_{\rm e}$, relaxing the friction offered by the foot. This effect eventually balances the capillary force, leading to a plateau-like 
behavior.  
Such remarkable surface specificity has been observed in experiments.
In Ref.~~\cite{Blake-AdvCollIntSci-2002}, water films were forced on surfactant-treated, dry aqueous gelatin surfaces, where $U_{\rm adv}$ was found to increase on hydrophilic surfaces to reach a maximum at $\theta_{\rm e} \approx 60-80^\circ$. 
In Ref.~\cite{Duez01}, a plateau for the splashing speed of glass beads treated with a hydrogen peroxide--sulphuric acid solution (hydrophilic beads) and a rapid decay for those grafted with silane chains (hydrophobic beads) was reported. 
Such a decay on hydrophobic surfaces is due to the absence of the liquid foot. Instead,  
the viscous friction is dominated by the gas wedge in front of the contact line, which becomes
increasingly narrow as $\theta \rightarrow 180^\circ$. 
In agreement with a previous analysis~\cite{Duez01}, in the hydrophobic limit Eq.~(\ref{eq:ForceBalance})  reduces to 
\begin{equation}
\label{eq:Lyderic}
(\pi-\theta_{\rm e})^2 -(\pi-\theta)^2 \simeq c_{\rm L}Ca +  \eta L(\pi-\theta)^{-1}Ca,
\end{equation}
where $L\sim {\cal O}(10)$ is a numerical pre-factor that depends on the scale separation, $\xi/h_{\rm o}$.     
In this regime the critical advancing speed obeys $ \eta_{\rm L}U_{\rm adv} \gamma \sim (\pi-\theta_{\rm e})^3$, as indicated by 
the scaling of the capillary (l.h.s.) and friction (r.h.s.) terms in Eq.~(\ref{eq:Lyderic}).

The viscosity contrast between the fluids also has a significant influence on fluid entrainment.
%
%
Here we focus on the effect of the solid, which we illustrate in the inset of Fig.~\ref{fig:CriticalCa}(b).    
While the critical point is shifted to a lower scale as $\eta$ is increased (reducing the gap between the advancing and receding critical velocities), 
the plateau-like behavior in hydrophilic substrates, due to the insensitivity of the advancing front to the foot shape at high $w_{\rm m}$, can persist to these variations.  
The critical dynamic angle, on the other hand, approaches 180$^\circ$ for decreasing viscosity of the displaced fluid, as has been shown 
numerically for plunging plate geometries~\cite{Marchand-PhysRevLett-2012}.
These results illustrate that the advancing configuration is distinct to the receding one due to the coupling between wetting, expressed in the shape of the profile, and the 
fluid viscosities.

{\it Conclusions.--} 
We have analyzed the impact of wetting on the onset of fluid entrainment on a forced  advancing  fluid front. We have put forward a general theoretical framework to identify
the morphological origin of fluid entrainment and the asymmetric role played by the affinity of the solid substrate to the advancing front. 
Our theory highlights the relevance of the fluid foot that the forcing liquid develops when the front advances on hydrophilic surfaces. The change in structure of the fluid front close to the solid, 
and the relative viscosity between the moving fluids are then shown to be responsible for the  asymmetry in the entrainment speed between advancing and receding configurations.  

The rich variety of wetting-entrainment regimes found highlights the important role of surface properties, characterized by the contact line friction coefficient $w_{\rm m}$. Recent developments
in atomic force microscopy~(AFM)~\cite{Xiong-PRE-2009} can probe forces associated to contact line dynamics~\footnote{P. Tong, private communication}, opening the possibility of quantifying 
the friction of the advancing contact line. 
The magnitude of the forces exerted on the probe support the feasibility of experiments that measure force profiles
as a function of the immersion speed of a dipping probe. 
Neglecting buoyancy forces, the force exerted by the AFM probe of diameter $d$ is balanced by the vertical component of the interfacial tension $T \approx \pi d \gamma \cos\theta_{\rm m} $, and 
by the viscous stress acting over the probe's surface.
For a $1 $~$\mu$m-wide probe~\cite{Xiong-PRE-2009}, the order of magnitude of these forces (below the transition), is set by $T(\theta_{\rm m} = \pi) = \pi d  \gamma$, and 
is of tens to hundreds of nano Newtons.
A detailed observation of the entrainment in the microfluidic regime is also possible, by using laser scanning microscopy 
of microfluidic experiments of demixed colloid-polymer mixtures of ultra-low surface tension~\cite{Aarts01}. 
Such experiments can be used to gain insight on the nature of the dynamic transition associated to entrainment and help in manufacturing  substrates with  well defined wetting properties to manipulate 
and better control the flow of liquid fronts at small scales. 

We thank Dirk Aarts, Penger Tong and Luis V\'azquez for enlightening discussions on the experimental applicability of
our theory, and Andrew Clark and Chris Lenn for useful discussions.  
RLA acknowledges support from Marie Curie Actions 
(FP7-PEOPLE-IEF-2010 no. 273406). AHM acknowledges partial financial support from MICINN (Spain) under project FIS2009-192964-
C05-02 and DURSI project SGR2009-00014.
IP acknowledges financial support from MINECO  (Spain) and DURSI under projects  FIS2011-
22603  and 2009SGR-634, respectively.


%

\end{document}